\documentclass[preprint,12pt]{elsarticle}

\usepackage{float}

\journal{Journal of Instrumentation}

\biboptions{sort&compress}

\begin{document}

\begin{frontmatter}

\title{Using Mineral Oil to Improve the Performance of Multi$-$crystal Detectors for Dark Matter Searching}





\renewcommand{\thefootnote}{\fnsymbol{footnote}}
\author{
J.C.~Liu$^{a}$,
C.~Guo$^{a,}\footnote{Corresponding author. Tel:~+86-1088236256. E-mail address: guocong@ihep.ac.cn (C.~Guo).}$,
Z.Y.~Yu$^{a}$,
M.Y.~Guan$^{a}$,
Z.M.~Wang$^{a}$,
X.H.Ma$^{a}$,
C.G.~Yang$^{a}$,
P.~Zhang$^{a}$,
C.J.~Dai$^{a}$,
W.L.~Zhong$^{a}$,
Z.H.~Li$^{a}$,
Y.P.~Zhang$^{a}$,
C.C.Zhang$^{a,b}$,
Y.T.~Wei$^{a,b}$,
W.X.~Xiong$^{a,b}$,
H.Q.~Zhang$^{a,b}$,
}
\address{
${^a}$Key Laboratory of Particle Astrophysics, Institute of High Energy Physics, Chinese Academy of Science,Beijing, China\\
${^b}$ School of Physics, University of Chinese Academy of Science, Beijing, China \\
}

\begin{abstract}
The inorganic crystals have been widely used for dark matter direct searching for many decades. However, limited by the crystal growth technique, a lot of small crystals have to be used together for large target mass, which results in a degradation of light collection efficiency. An experiment was built up to study the degradation, and the method of soaking crystals into mineral oil to improve the efficiency as well as reduce the interface effect were proposed and validated. Good data and MC agreements were achieved in the experiment.
\end{abstract}


\end{frontmatter}


\section{Introduction}

The Weakly Interacting Massive Particle(WIMP) is a popular candidate for dark matter, which is observed by its gravitational influence on nearby matter\cite{HPA6_110_127,APJ159_379_403,APJ238_471,PhD,MNRAS311_441_447}. The Planck experiment shows that $\sim$84.5\% of the universe's matter is dark matter\cite{AA517_A16}. Meanwhile, the dark matter direct searching experiments, seeking the evidence for low energy($\le$ 50 keV) nuclear recoils caused by elastic scattering of WIMPs, are currently being carried out by several groups with different targets, including  scintillator crystals\cite{Eur76_25,PLB616_17}, liquid xenon\cite{PRL118_021303,PRL118_071301,arxiv}, liquid argon\cite{PRD93_081101}, high purity semi-conductors\cite{Front8_412,PRL116_071301}, etc.

Crystals, one of the earliest detectors for dark matter direct detection, are still widely used today. Neutron beam tests for crystals indicate that crystal could be or have the potential to be good targets for dark matter searching experiments\cite{NIMA491_460,NIMA818_38_44,NIMA833_49_53}. The Dark Matter Experiment(DAMA) are running with  $\sim$250~kg highly radiopure NaI(Tl) crystals and reported the dark matter annual modulation signature at 9.3~$\sigma$\cite{EPJ136_05001}. The Korea Invisible Matter Search(KIMS) experiment, running with 103.4~kg CsI(Tl) crystals, obtained very competitive results especially in the spin-dependent(SD) detection\cite{PRL108_181301}. However, the conventional design, using an outer copper encapsulation, limit the sensitivity because of its relatively poor light collection, which is caused largely by the light loss at the many components, boundaries and surfaces of crystals and PMTs. Any improvement of light collect efficiency is crucial to detector sensitivity because the energy spectrum due to WIMPs is expected to fall faster than the background so that the low energy data, near threshold, are the most significant in determining sensitivity\cite{AP6_87}. On the other hand, limited by the crystal growth technique, ton-scale crystals are difficult to manufacture and the interface effect prevent the multi-crystals to perform as good as a single crystal. Thus the detectors have to be made in multi-arrays.

The project of using mineral oil to reduce the interface effect and improve the light yield of crystals was proposed and a prototype, as well as Monte Carlo(MC) simulation program, were constructed to verify the assumption. In this paper, the interface effect reduction factor for combined crystal and the light collection efficiency improvement for multi-crystal detector from both MC simulation and experiment, which are consistent with each other, are presented.

\section{Experimental setup}

\begin{figure}[H]
\centering
\includegraphics[width=6cm,height=6cm]{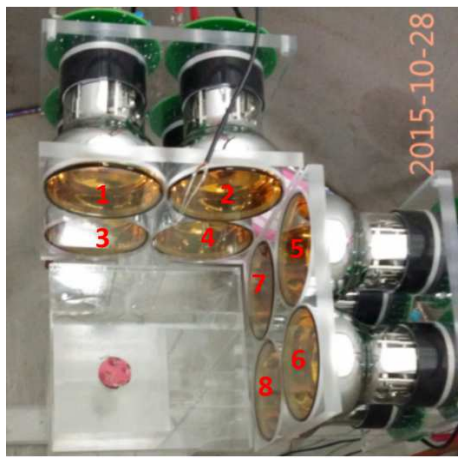}
\caption{Photo of the experiment, eight PMTs facing a cubic crystal. }
\label{fig.protype_1}
\end{figure}

A prototype detector, based on CsI(Tl) crystals, is constructed to study the degradation of light collect efficiency and to validate the proposal of using mineral oil to improve the light collection efficiency as well as to reduce the interface effect. Since the emission peak of CsI(Tl) crystal does not match with normal bi-alkali Photon Multiplier Tube(PMT), eight R6233-100 PMTs from Hamamatsu company are used\cite{CsI_datasheet,R6233_100_datasheet}. Figure~\ref{fig.protype_1} shows a photo of the experimental setup. Five polished CsI(Tl) crystals, two of them are 7.5$\times$7.5$\times$7.5~cm$^3$ in cubic shape and the other three are 7.5$\times$7.5$\times$15~cm$^3$ in rectangular shape, are used in the experiment. The CsI(Tl) crystals, with 0.02\% Tl doping concentration, are produced by the Beijing Glass Research Institute.The whole system is located at the center of a stainless steel cylinder of which the height and diameter are both 1~m.  When the detectors are tested in the air, the crystal-PMT surfaces are coated with mineral oil to reduce the influence of the interfaces between them as well as to make them stick together. While they are tested in mineral oil, the stainless steel cylinder is fully filled.

\begin{figure}[htb]
\centering
\includegraphics[height=6.5cm]{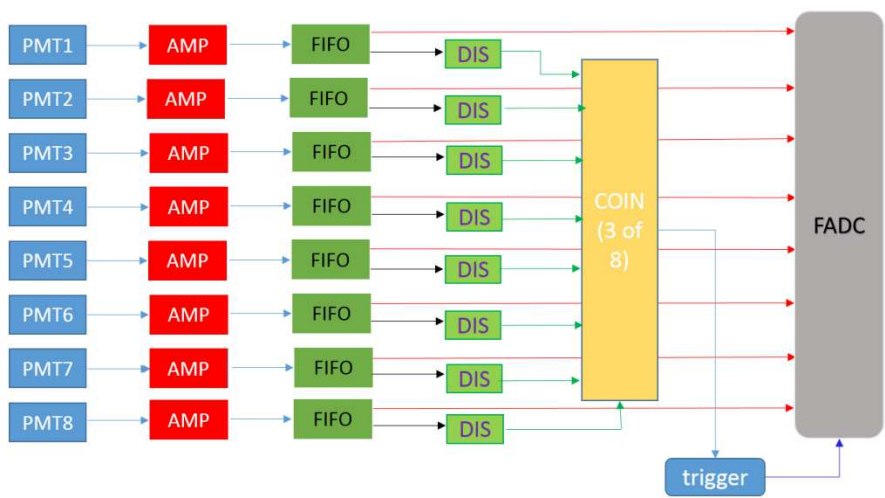}
\caption{Readout diagram of the protype.}
\label{fig.trigger}
\end{figure}

Figure~\ref{fig.trigger} shows the electronics scheme of the experiment. In total there are eight channels from the eight PMTs. The data are recorded by the Flash Analog to Digital Convertor~(FADC CAEN V1792A, 1~GHz sampling frequency, 2.5~$\mu$s readout window).  At the overall supply voltage, the gain of the PMTs are $\sim$2$\times$10$^{5}$ thus single photoelectron~(p.e.) peak can not be observed. A sixteen channel fast amplifier(AMP CAEN N979, 10 times fixed gain)~\cite{caen} is used to amplify the signals after the quad linear Fan in/Fan out~(CAEN N625, 100MHz bandwidth, $\pm$1.6V maximum input amplitude). In order to reduce the background caused by the random coincidence, the FADC trigger is generated by the coincidence of three or more PMTs, where the single channel threshold is about 3~p.e..

\section{Experimental and MC simulation results}
\begin{figure}[H]
\centering
\includegraphics[width=6cm,height=9cm]{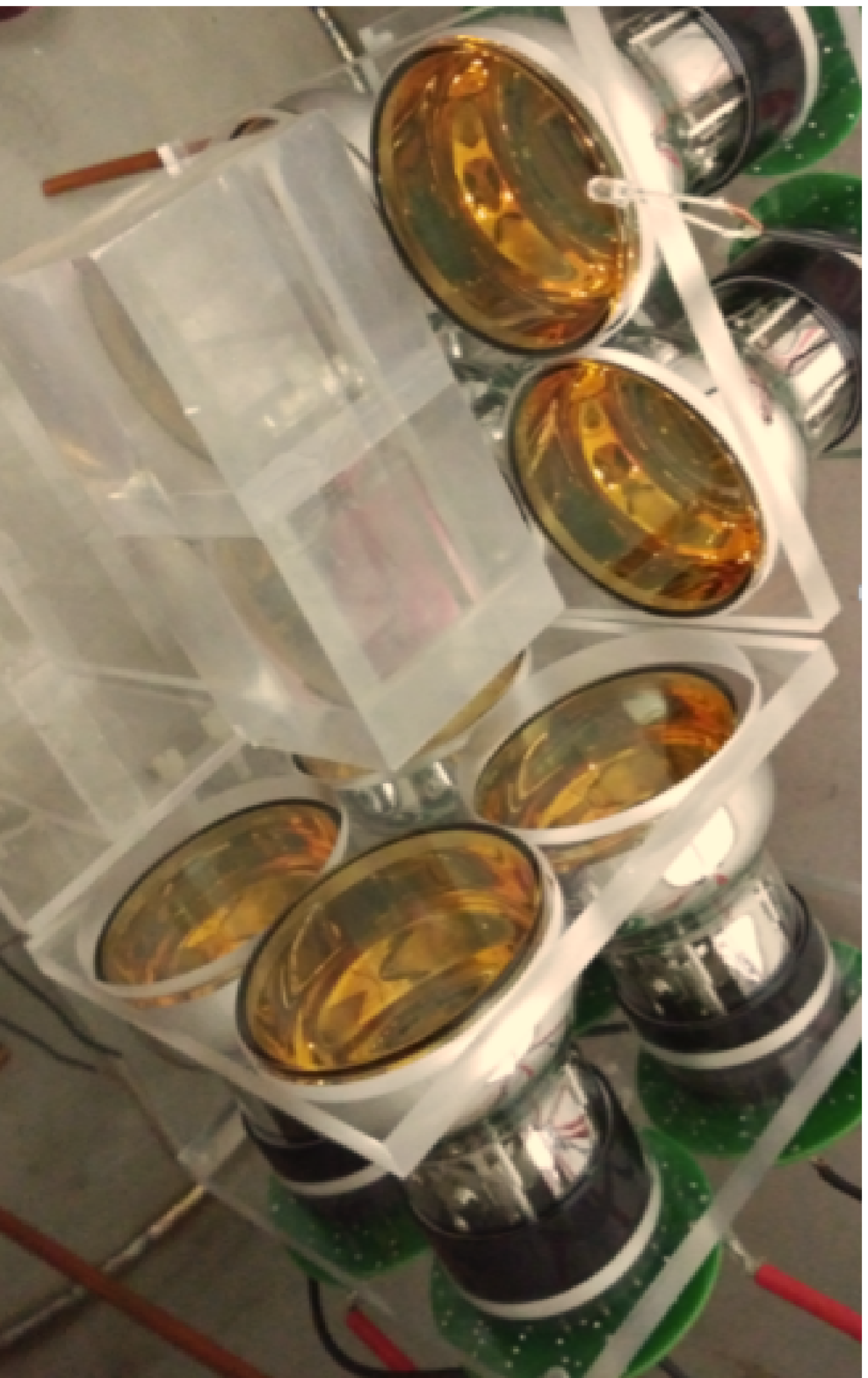}
\includegraphics[width=6cm,height=9cm]{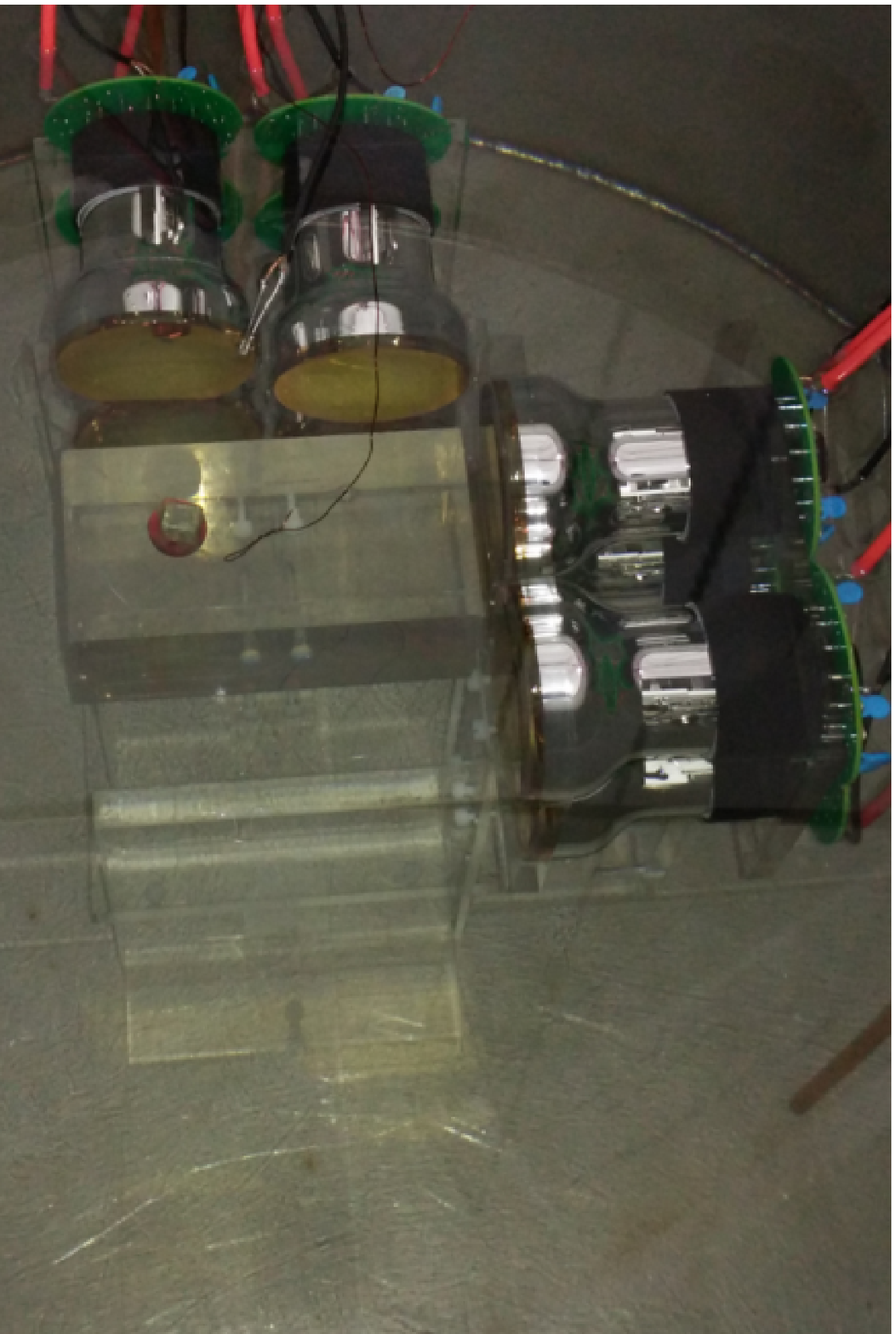}
\caption{Left: Combined crystal in air. Right: One monolithic rectangular crystal in mineral oil.}
\label{fig.interface_effect}
\end{figure}
According to Ref.~\cite{PLB473_330_336}, the mineral oil is a good candidate for being the matching liquid between crystals because of its similar refractive index and long term compatibility with CsI(Tl). In this section, the experimental procedures are described in detail. The full energy peaks irradiated by $^{137}Cs$ are used to study the interface effect reduction and light collection efficiency improvement. The events shown in the energy spectra with higher energy than the full energy peak of $^{137}Cs$ come mainly from the environmental radiation and cosmic ray. The gap between the crystals are $\sim$0.1~mm during the experiment and during the simulation it is set to 0.1~mm, while the gap between the crystal and the PMT surfaces is $\sim$0.0~m.

\subsection{Interface effect}
\begin{figure}[htb]
\centering
\includegraphics[height=6.5cm]{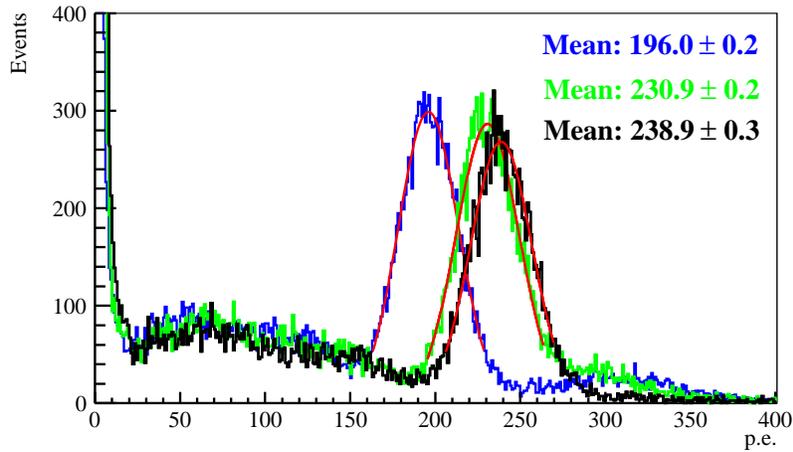}
\caption{Energy spectra from PMT7, the blue line is for the combined crystals in air, the green line is for the combined crystal in mineral oil and the black line is for the monolithic crystal in mineral oil. }
\label{fig.spectrum_2_1}
\end{figure}
In order to study the performance of mineral oil in the reduction of interface effect for the crystal detector, one combined rectangular crystal and one monolithic rectangular crystal are placed in the stainless cylindrical vessel respectively. As is shown in fig.\ref{fig.interface_effect}, the combined crystal, which consist of two cubic crystals, is with the same size as the monolithic one. During the tests, the trigger is as fig.\ref{fig.trigger} shows. A $^{137}Cs$ $\gamma$ source is placed at the surface of the left crystal so that the p.e. collected by PMT7 can be used to study the interface effect between the two cubic crystals. By comparing the energy spectra of PMT7 in three different cases, the effect of mineral oil in the reduction of interface effect can be gotten. The $\gamma$ source has been kept in the same place during the tests. Fig.\ref{fig.spectrum_2_1} presents the energy spectra as well as fitting results of combined crystal in air, combined crystal in mineral oil and monolithic crystal in oil. A factor D is defined as equ.\ref{Eq:differ} to describe the reduction of interface effect.
\begin{equation}
D = \frac{E_2-E_1}{E_2}
\label{Eq:differ}
\end{equation}
where $E_{2}$ is the $^{137}Cs$ full energy peak of the monolithic rectangular crystal in mineral oil and $E_{1}$ is the $^{137}Cs$ full energy peak of the combined crystal in air or in mineral oil respectively. As can be seen from the energy spectra, the difference between combined crystal and the monolithic one has been reduced from 17.92\%$\pm$0.03\% to 3.35\%$\pm$0.01\% after been soaked into the mineral oil, which means that the mineral oil can sufficiently reduce the interface effect of the combined crystal. The errors of D are statistical only.

In order to obtain the expected measured spectra of the crystals, which are irradiated by $^{137}Cs$, a MC simulation with Geant4 is used. The geometry constructed by the program is shown in fig.5 and the effects of crystal geometry, $\gamma$ source geometry, electronic response are included in the simulation. The detail information about the CsI(Tl) crystal can be found in ref\cite{CsI_datasheet} and detail information about the PMTs can be found in ref\cite{R6233_100_datasheet}. The simulated spectra with fitting results are shown in fig.~\ref{fig.simulation_2_1}, which are basically consistent with the data.

\begin{figure}[htb]
\centering
\includegraphics[height=6.5cm]{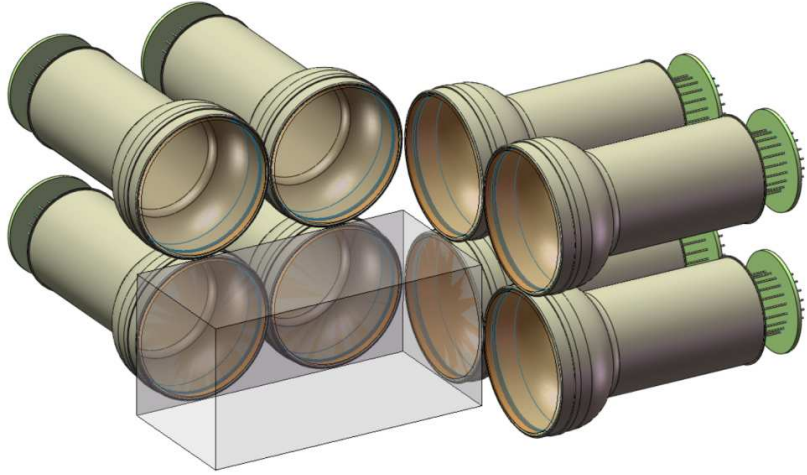}
\caption{The geometry constructed by the program. In the graph, one 7.5$\times$7.5$\times$15 $cm^3$ rectangular crystal is placed.}
\label{fig.simulation}
\end{figure}

\begin{figure}[htb]
\centering
\includegraphics[height=6.5cm]{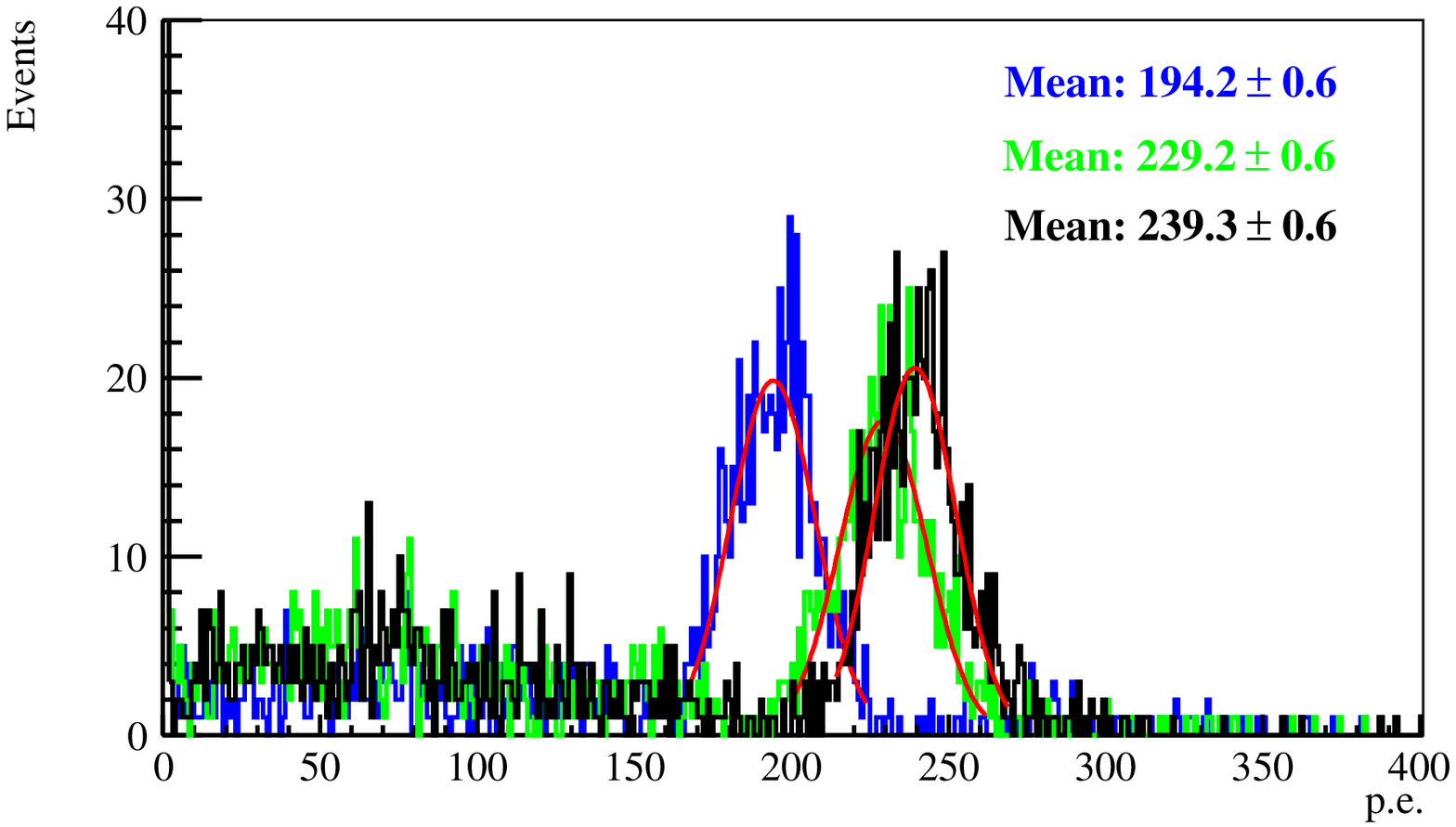}
\caption{MC result of energy spectrum from PMT7, the blue line is for two cubic crystals in air, the green line is for two cubic crystals in mineral oil and the black line is for the rectangular one in mineral oil.}
\label{fig.simulation_2_1}
\end{figure}

\par
It is known to all that when the photons reach a boundary between different materials with different refractive indices, the photons will be partially refracted and partially reflected. However, if the photons propagate from optically denser medium to optically thinner media and the angle of incidence is greater than the critical angle, which is determined by equ.\ref{Eq:theta}, the photons will be totally reflected back.

\begin{equation}
\theta = arcsin(\frac{n_2}{n_1})
\label{Eq:theta}
\end{equation}

where $\theta$ is the critical angle, $n_{1}$ is the refractive index of optically denser medium and $n_{2}$ is the refractive index of optically thinner medium. In this experiment, $n_{1}$ is the refractive index of CsI(Tl) crystal and $n_{2}$ is the refractive index of air or mineral oil respectively. Since the crystals do not have infinite attenuation length and the refractive index of mineral oil is higher than air, which will cause a greater critical angle and thus a longer average track length of photons in crystal, more photons will be absorbed by the crystals when they are put in the air. Thus the number of p.e. collected by PMT7 is larger for crystal been soaked in mineral oil than put in the air, which are consistent with the results shown in fig.~\ref{fig.spectrum_2_1} and fig.~\ref{fig.simulation_2_1}.

\subsection{Improvement of the collect efficiency}
\subsubsection{Result for current prototype and MC simulation }
\begin{figure}[H]
\centering
\includegraphics[width=6cm,height=6.5cm]{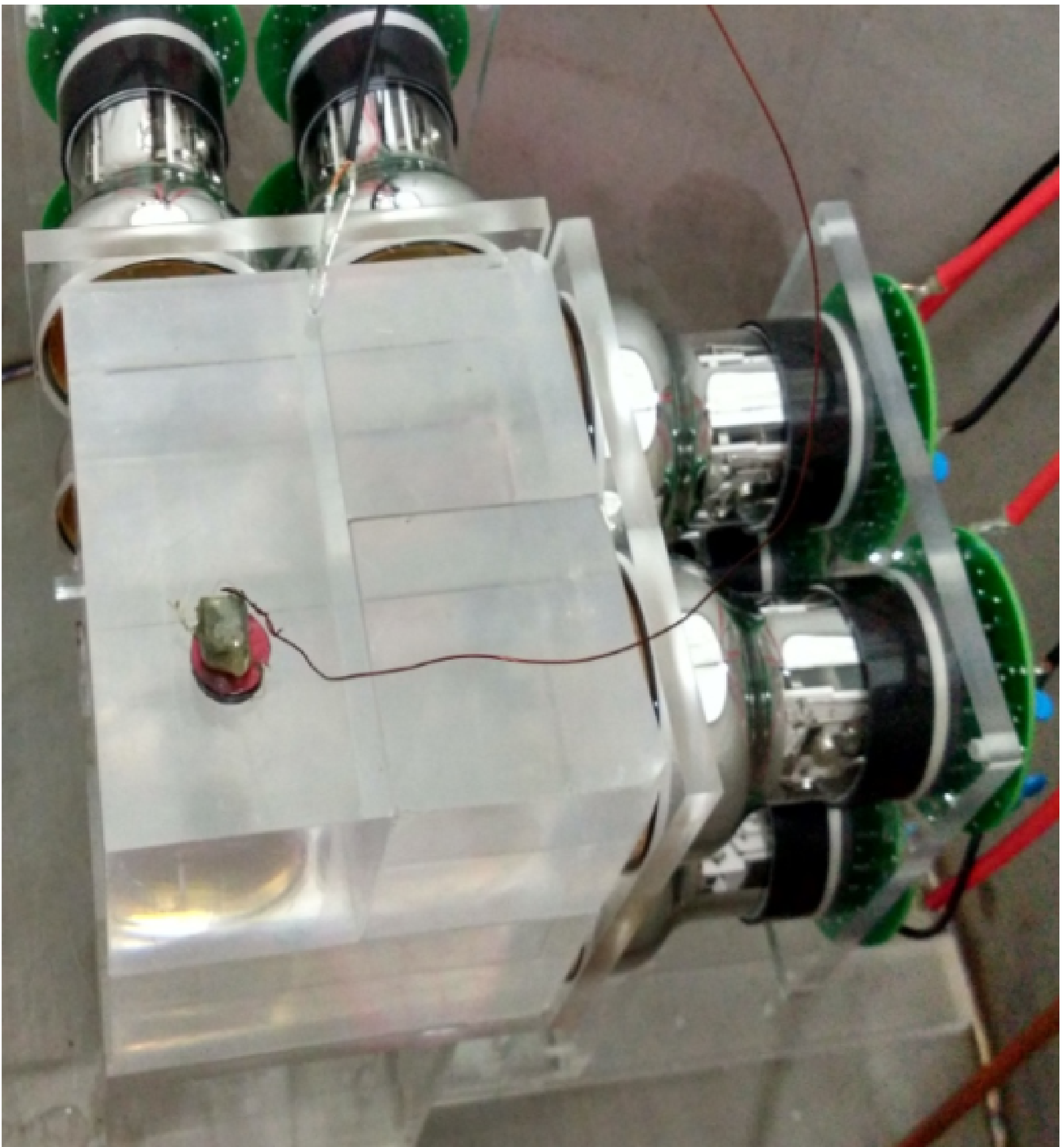}
\includegraphics[width=6cm,height=6.5cm]{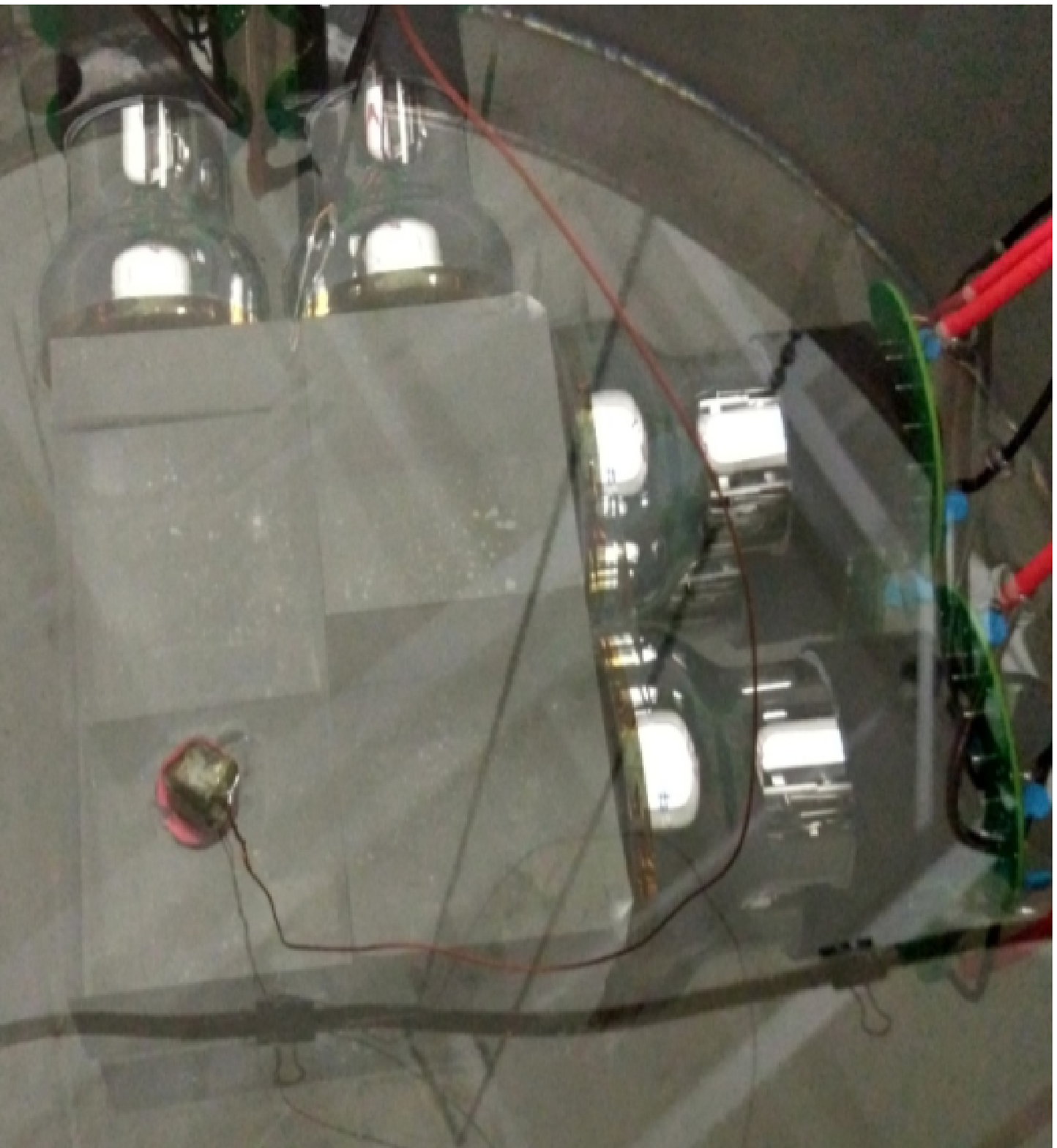}
\caption{The multi-crystal detector. Left: Detector in the air; Right: Detector soaked in the mineral oil. }
\label{fig.prototype_all}
\end{figure}

\begin{figure}[H]
\centering
\includegraphics[height=6.5cm]{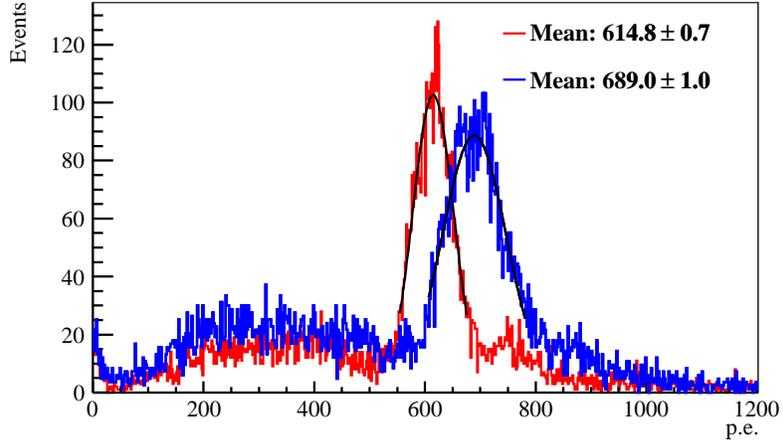}
\caption{The number of p.e. collected by all the PMTs. The red line is the result of the multi-crystal detector in air and the blue line is the result of the same detector in mineral oil.}
\label{fig.spectrum_all_2}
\end{figure}

\begin{figure}[H]
\centering
\includegraphics[height=6.5cm]{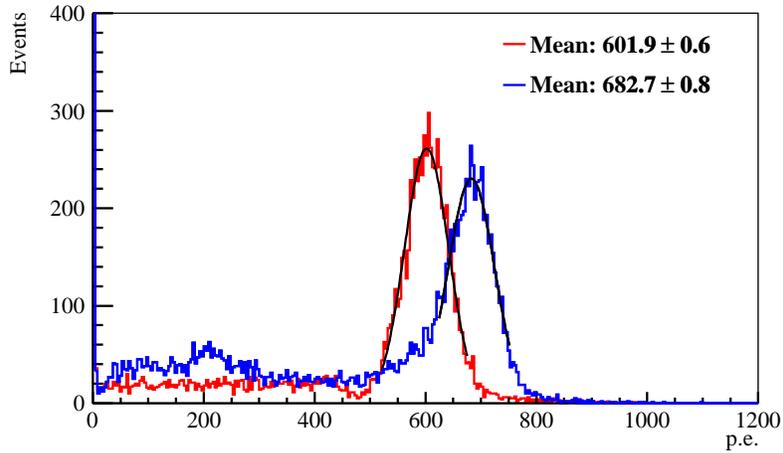}
\caption{Simulated results of the total p.e. collected by all the PMTs. The red line is the result of multi-crystal detector in air and the blue line is the result of same detector in mineral oil.}
\label{fig.simulation_all}
\end{figure}
To test the effect of mineral oil in improving the light collect efficiency, a prototype(fig.\ref{fig.prototype_all}) which consist of a 15$\times$15$\times$15~$cm^{3}$ cubic multi-crystal detector and eight PMTs is constructed. In order to make the photons propagate as many interfaces as possible, the $^{137}Cs$ $\gamma$ source is placed at the far end of the diagonal. As can be seen from the spectra shown in fig.\ref{fig.spectrum_all_2}, the light yield of the prototype has been increased 11.73\%$\pm$0.02\% after been soaked into the mineral oil.  The simulated result, 12.01\%$\pm$ 0.09\%, is shown in fig.\ref{fig.simulation_all}, which are basically consistent with the data. The errors are also statistical only.

\subsubsection{MC simulation result of future possible multi-crystal detector}
\begin{figure}[H]
\centering
\includegraphics[width=6cm,height=6.5cm]{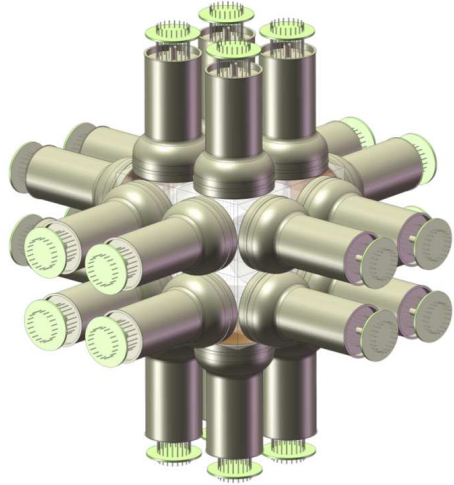}
\caption{The geometry of the multi-crystal detector.}
\label{fig.all_all}
\end{figure}

\begin{figure}[htb]
\centering
\includegraphics[height=6.5cm]{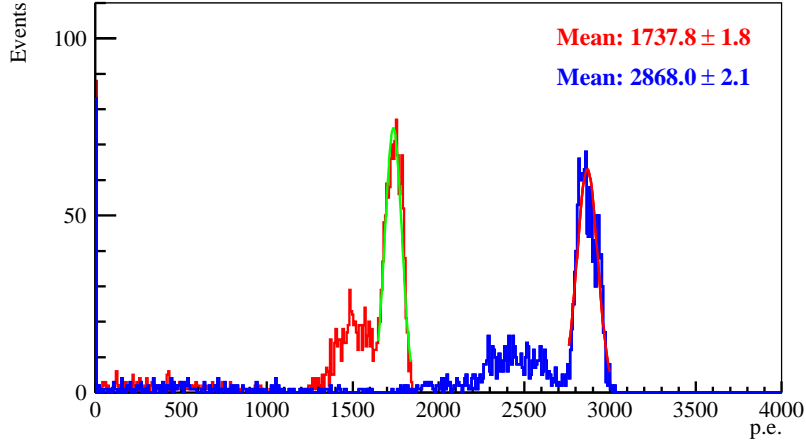}
\caption{Simulated results of the total p.e. for multi-crystal detector. The red line is the multi-crystal detector in air and the blue line is the detector in mineral oil.}
\label{fig.16PMT_64Cr_spectrum}
\end{figure}

It is known to all that the smaller pieces of crystals are more likely to have better quality as well as relatively lower cost. In addition, dark matter searching experiments are more inclined to use identical crystals to build crystal-array detectors. Thus the geometry shown in fig.~\ref{fig.all_all} is constructed to simulate the effect of mineral oil in improving the light collecting efficiency. Sixty four 1.875$\times$1.875$\times$ 1.875~$cm^{3}$ cubic crystals are used to form a 15$\times$15$\times$15~$cm^{3}$ cubic multi-crystal detector, and sixteen PMTs are placed at the six surfaces for readout. A dot $^{137}Cs$ $\gamma$ source is placed in the center of the detector to irradiate the crystals so that the p.e. collected by all the PMTs can be used to describe the light yield improvement.  Fig.~\ref{fig.16PMT_64Cr_spectrum} shows the simulated result, the red line are the total p.e. collected by all the PMTs for air coupling and blue line is for mineral oil coupling. As can be seen from the plots, the light collect efficiency has been increased 65.04\%$\pm$0.08\% after been soaked into the mineral oil. The error is also statistical only.

\section{conclusion}
The effect of mineral oil in the reduction of interface effect for combined crystal, as well as the improvement of light collecting efficiency for multi-crystal detector, has been studied with CsI(Tl) crystals. The interface effect reduction factor and the light yield improvement, calculated with $^{137}Cs$ $\gamma$ source data, have been reported in this paper. A 15${\times}$15${\times}$15~$cm^{3}$ cubic multi-crystal detector made of five crystals or made by sixty four cubic crystals, with eight or sixteen PMTs readout will lead to a $\sim$12\% or ${\sim}$65\% improvement, respectively, when mineral oil fills the gaps between the surfaces of different crystals, replacing the air.
\par
The experimental results, which are consistent with the MC simulation results, validate the proposal of using the mineral oil to reduce the interface effect of combined crystal and to increase the light yield of a multi-crystal detector. The MC simulation results for future possible multi-crystal detector implied that the effect of mineral oil in improving the light yield would be more obvious if more crystals are used to form the multi-crystal detector.

\section{Acknowledgements}
This work is supported by the Ministry of Science and Technology of the People's Republic of China (No. 2010CB833003). Many thanks to Y.S.Lv, H.Q.Lu and J.L.Xu of IHEP for their help during the experiment.

\end{document}